\documentclass[10pt,preprint2,amssymb]{aastex6}

\usepackage{graphicx}
\renewcommand{\vec}[1]{\boldsymbol{#1}}
\usepackage{amsmath}
\def\pop{PhPl}

\shorttitle{Compressible plasmoid-dominated reconnection}
\shortauthors{Zenitani \& Miyoshi}

\begin{document}

\title{Plasmoid-dominated turbulent reconnection in a low $\beta$ plasma}

\author{Seiji Zenitani\altaffilmark{1}}
\altaffiltext{1}{Research Center for Urban Safety and Security, Kobe University, 1-1 Rokkodai-cho, Nada-ku, Kobe 657-8501, Japan}
\author{Takahiro Miyoshi\altaffilmark{2}}
\altaffiltext{2}{Department of Physical Science, Graduate School of Science, Hiroshima University, Higashi-Hiroshima 739-8526, Japan}

\begin{abstract}
Properties of plasmoid-dominated turbulent reconnection
in a low-$\beta$ background plasma are investigated
by resistive magnetohydrodynamic (MHD) simulations.
In the $\beta_{\rm in} < 1$ regime,
where $\beta_{\rm in}$ is plasma $\beta$ in the inflow region,
the reconnection site is dominated by shocks and shock-related structures and plasma compression is significant. 
The effective reconnection rate increases from $0.01$ to $0.02$ as $\beta_{\rm in}$ decreases.
We hypothesize that
plasma compression allows faster reconnection rate,
and then
we estimate a speed-up factor, based on a compressible MHD theory.
We validate our prediction by a series of MHD simulations. 
These results suggest that
the plasmoid-dominated reconnection can be twice faster than expected
in the $\beta \ll 1$ environment in a solar corona.
\end{abstract}

\section{Introduction}

Magnetic reconnection is an important process
to change magnetic topology and
to release magnetic energy in solar, space, and astrophysical plasmas. 
In magnetohydrodynamics (MHD), reconnection has long been discussed
by Sweet--Parker \citep{sweet,parker} and Petschek models \citep{petschek}. 
In Sweet--Parker theory, the reconnection rate is
given by $\propto S^{-1/2}$,
where $S$ is a system-size parameter called the Lundquist number.
A problem was that
the Sweet--Parker rate is unable to explain reconnection events in the Universe.
On the other hand,
the Petschek reconnection achieves fast reconnection rate, but
it assumes a localized diffusion region near the X-line. 
It requires ad-hoc prescriptions
such as spatially-localized or parameter-dependent resistivities \citep{scho89,ugai92}.

Earlier theorists envisioned that
a laminar Sweet--Parker layer may break up into multiple magnetic islands (plasmoids) \citep{biskamp86,tajima97,shibata01}.
A subsequent theory has shown that
the reconnection layer will not be laminar,
because a tearing-type instability grows in the Sweet--Parker layer within an Alfv\'{e}n-transit time in high-$S$ systems \citep{loureiro07}.
As a consequence, the Sweet--Parker reconnection switches to plasmoid-dominated turbulent reconnection \citep{lapenta08}
at high Lundquist-number regime of $S \gtrsim S_{\rm c}$ \citep{bhattacharjee09},
where the critical Lundquist number $S_{\rm c}$ is on an order of $\mathcal{O}(10^4)$.
Importantly, the reconnection rate during the plasmoid-dominated stage remains constant $\sim 0.01$,
regardless of $S$ and other parameters. 
Then the MHD reconnection rate can be moderately fast in the high-$S$ regime, without the help of Petschek mechanism. 
The tearing-type instability and/or the plasmoid-dominated reconnection is popularly called the {\itshape plasmoid instability}, and it has been actively studied in the last decade \citep{huang13,loureiro16}. 
The transition from the Sweet--Parker reconnection to the plasmoid-dominated reconnection is analogous to
the one from a laminar flow to a turbulent flow in fluid dynamics. 
In both cases, dimensionless system-size parameters,
the Ludquist number or the Reynolds number, characterize the system. 
Similarly, other parameters in fluid dynamics would be applicable to the reconnection problem. 

Plasma $\beta$ ($\equiv {p_{\rm gas}}/{p_{\rm mag}}$)
in the inflow region, $\beta_{\rm in}$, 
is a key parameter in the reconnection system. 
Since the speed of a reconnection jet reaches
the inflow Alfv\'{e}n speed $c_{\rm A,in}=|B_{\rm in}| / \rho_{\rm in}^{1/2}$,
the inflow plasma $\beta$ ($\beta_{\rm in}$) determines the sonic Mach number of the reconnection jet,
$\mathcal{M}_s \equiv c_{\rm A,in}/c_{\rm s,in} = (\frac{1}{2}\gamma \beta_{\rm in} )^{-1/2}$. 
In analogy with fluid dynamics,
the reconnection system should involve compressible effects
such as shock-formation and plasma compression
in the low $\beta_{\rm in}$ (high $\mathcal{M}_s$) regime.
However, even though plasma $\beta$ is extremely low ($\beta \ll 1$) around reconnection sites in a solar corona \citep{gary01},
many studies on plasmoid-dominated reconnection explore the $\beta_{\rm in} > 1$ regime.
Only \citet{ni12,ni13} and \citet{baty14} investigated
the influence of plasma $\beta_{\rm in}$ on the critical Lundquist number $S_{\rm c}$ that determines the onset of the turbulent state. 
Many properties of the plasmoid-dominated reconnection in the low $\beta_{\rm in}$ regime remain unclear.

\begin{figure*}[htbp]
\centering
\includegraphics[width={0.91\textwidth}]{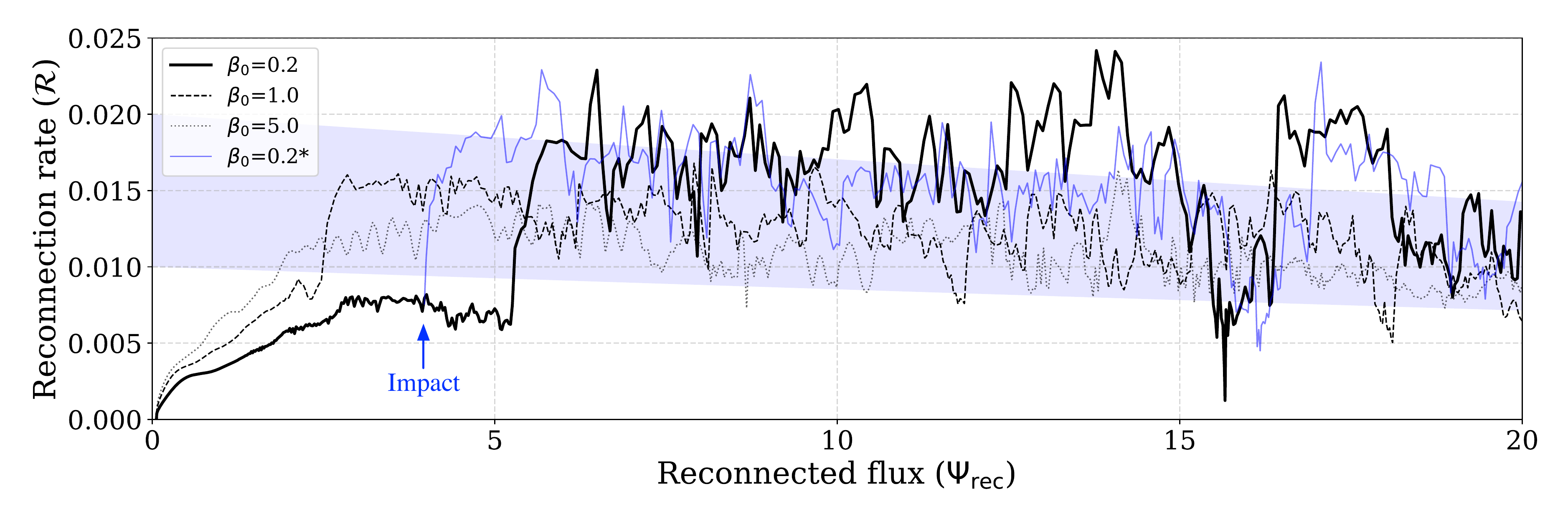}
\caption{
Time evolution of the reconnection rate $\mathcal{R}$
as a function of the reconnected flux $\Psi_{\rm rec}$.
The blue line indicates the rate of an additional run for $\beta_0=0.2$,
in which an artificial impact is imposed at $t=1000$.
The blue shadow indicates a region of $0.01<\mathcal{\bar{R}}<0.02$,
where $\mathcal{\bar{R}}$ is the normalized rate (see the text).
\label{fig:rate}}
\end{figure*}

This Letter explores basic properties of
plasmoid-dominated reconnection in the $\beta_{\rm in} < 1$ regime.
By means of large-scale resistive MHD simulation,
we investigate the influence of compressible parameters
such as the inflow $\beta$ ($\beta_{\rm in}$) and the specific heat ratio $\gamma$.

\section{Numerical setup}
\label{sec:setup}

We use a finite-volume MHD code, {\texttt OpenMHD} \citep{zeni15a,openmhd}. 
It employs an HLLD Riemann solver \citep{miyoshi05}
to deal with shocks and discontinuities. 
Simulations are carried out in the $x$-$y$ plane.
The initial magnetic field, velocity, density, and pressure are given by
$\vec{B} = B_0 \tanh(y/l) \vec{\hat{x}}$, $\vec{v}=0$,
$\rho(y) = \rho_0 [ 1 + \cosh^{-2}(y/l)/ \beta_0]$, and
$p(y) = 0.5 \beta_0 \rho(y)$.
Here $l=1$ is a current-sheet thickness,
$B_0=1$, $\rho_0=1$, and
$\beta_0$ is the initial plasma $\beta$ in the inflow (background) region. 
We set $\beta_0 = 0.2, 1.0$, and $5.0$ for our main runs.
The adiabatic index is set to $\gamma=5/3$.
The other symbols have their standard meanings. 
The parameters are normalized such that
the inflow Alfv\'{e}n velocity is $c_{\rm A0} = |B_0|/\sqrt{\rho_0}=1$. 
The time is normalized by the Alfv\'{e}n crossing time of the current sheet, $l/c_{\rm A0}=1$. 
The domain size is $[-L_x,L_x] \times [0,L_y]$
with $L_x=500$ and $L_y=100$. 
It is resolved by $30000 \times 3000$ grid points.
We set periodic boundaries at $x=\pm L_x$,
a reflecting boundary at $y=L_y$,
and a mirror boundary at $y=0$. 
The resistivity is fixed to $\eta =10^{-3}$. This corresponds to a magnetic Reynolds number of $R_m \equiv c_{\rm A0} l / \eta = 10^3$. Using a quarter length of the system $L_x/2$, a typical Lundquist number is $S_L \equiv c_{\rm A0} (L_x/2) / \eta = 2.5 \times 10^5$.
Magnetic reconnection is triggered by a localized perturbation, $\delta A_z = 0.06 \exp[-(x^2+y^2)/4]$.

\begin{figure*}[htbp]
\centering
\includegraphics[width={0.91\textwidth}]{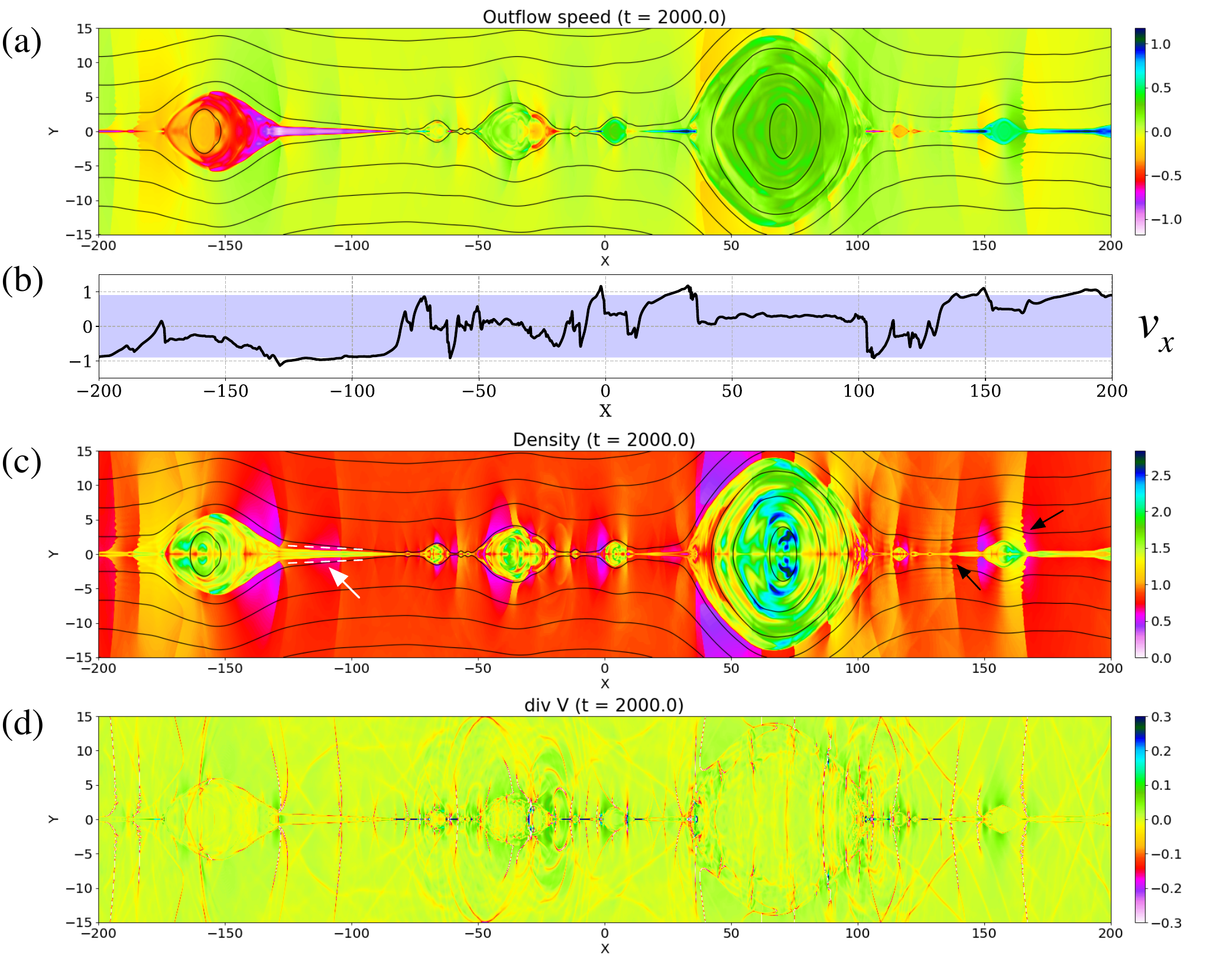}
\caption{
(a) Plasma outflow velocity $v_x$ at $t=2000$ and
(b) 1-D cut of $v_x$ at $y=0$,
(c) plasma density, and
(d) divergence ($\nabla\cdot\vec{v}$) of the $\beta_0=0.2$ run
at t=2000 are presented.
The blue shadow in (b) indicates the range $-c_{\rm A,in} < v_x < +c_{\rm A,in}$.
\label{fig:snapshot}}
\end{figure*}

\section{Results}
\label{sec:results}

The black lines in Fig.~\ref{fig:rate} show
the reconnection rate,
a flux transfer rate by the reconnection process,
as a function of a reconnected flux.
We calculate a flux function $\Psi(x) = \int_{-L_x}^{x} B_y(x) dx$ at y=0, and then
we estimate the reconnected flux
$\Psi_{\rm rec} \equiv \Psi_{\rm max} - \Psi_{\rm min}$ and
the reconnection rate
$\mathcal{R} = \frac{1}{c_{\rm A0}B_0}\partial_t \Psi_{\rm rec} = \partial_t \Psi_{\rm rec}$. 
We initially observe slow and laminar evolution. 
At some point ($\Psi_{\rm rec}=5.3$ for $\beta_0$=0.2), the system turns into a plasmoid-dominated turbulent stage.
As will be shown, small magnetic islands (plasmoids) are repeatedly generated, and the reconnection rate increases to $\mathcal{R}= 0.015$--$0.02$.
In all cases, the rate gradually decreases in time,
as the system consumes the magnetic flux,
$\Psi_0 = \int_0^{L_y}B_x {\mathrm d}y \big|_{t=0} = 99.2$.
We estimate
the magnetic field in the inflow region $B_{\rm in} \approx B_0 (1-\Psi_{\rm rec}/\Psi_0)$ and
the plasma density $\rho_{\rm in} \approx \rho_0 (1-\Psi_{\rm rec}/\Psi_0)$.
Then, since the rate is controlled by the inflow properties, $\mathcal{R} \propto c_{\rm A,in} B_{\rm in} = B_{\rm in}^2 / \rho_{\rm in}^{1/2}$,
we expect $\mathcal{R} \approx \mathcal{\bar{R}} (1-\Psi_{\rm rec}/\Psi_0)^{3/2}$, where $\mathcal{\bar{R}}$ is a normalized reconnection rate. 
The blue shadow in Fig.~\ref{fig:rate} indicates
$0.01 < \mathcal{\bar{R}} < 0.02$. 
One can see that
the normalized rates $\mathcal{\bar{R}}$ remain similar
in time-averaged sense for all cases. 
From $\partial_t\Psi_{\rm rec} = \mathcal{\bar{R}} (1-\Psi_{\rm rec}/\Psi_0)^{3/2}$,
we derive $\Psi_{\rm rec}(t) = \Psi_0 [1-(1+\mathcal{\bar{R}}t/2\Psi_0)^{-2}]$ and then
estimate an effective normalized rate
$\langle\mathcal{\bar{R}}\rangle$ between $7<{\Psi_{\rm rec}}<17$:
\begin{align}
\langle\mathcal{\bar{R}}\rangle =
\frac{2\Psi_0}{t_{\Psi_{\rm rec}=17}-t_{\Psi_{\rm rec}=7}}
\Big( \sqrt{\frac{\Psi_0}{\Psi_0-17}} - \sqrt{\frac{\Psi_0}{\Psi_0-7}} \Big)
.
\end{align}
We obtain
$\langle\mathcal{\bar{R}}\rangle = 0.0188, 0.0148,$ and $0.0130$
for $\beta_0=0.2, 1.0$, and $5.0$, respectively. 
Surprisingly, the reconnection rate during the plasmoid-dominated reconnection becomes higher for $\beta_{\rm in} < 1$.

Let us see visible signatures of the reconnection system in the $\beta_{\rm in} < 1 $ regime.
Fig.~\ref{fig:snapshot} displays various quantities of the $\beta_0=0.2$ run
in a central region ($x \in [-200, 200]$ and $y \in [-15,15]$) at $t=2000$ ($\Psi_{\rm rec}=18.0$).
We also present the bottom halves of $y<0$
to guide the reader's eyes.
As documented in many studies, the system is filled with a lot of plasmoids, repeatedly generated in inter-plasmoid layers.
The left-right asymmetry purely originates from numerical noise. 
The outflow velocity ($v_x$) is bounded by an estimated inflow Alfv\'{e}n velocity,
$|v_x| < c_{\rm A,in} \approx (B_0/\rho_0^{1/2}) (1-\Psi_{\rm rec}/\Psi_0)^{1/2} |_{\Psi_{\rm rec}=18} \approx 0.9 c_{\rm A0}$ (Figs.~\ref{fig:snapshot}(a) and \ref{fig:snapshot}(b)). 
The initial current sheet is completely ejected from the domain.

One of the clearest signatures is
the normal shocks around the reconnection sites,
as evident in the velocity/density jumps in Figs.~\ref{fig:snapshot}(a), \ref{fig:snapshot}(c), and
in the red regions in Fig.~\ref{fig:snapshot}(d). 
They are vertical slow shocks, generated by plasmoids \citep{zeni11a,tanuma07}.
They travel in the left and right directions across the reconnection sites.
The shocks travel even inside the plasmoids and intersect each others. 
One can see finger-like structures of shockfronts,
as indicated by the black arrows in Fig.~\ref{fig:snapshot}(c).
They are attributed to the corrugation instability \citep{stone95,zeni11a}. 
Due to these shocks, shock-shock interactions, and shock-related instabilities,
the system becomes highly complex in the $\beta_{\rm in} < 1$ regime. 
Also, we occasionally observe Petschek-like structures with bifurcated slow-shock layers,
in agreement with recent studies \citep{mei12,baty12,shibayama15}.
They are indicated by the white arrow in Fig.~\ref{fig:snapshot}(c). 
These Petschek-like structures are found behind outgoing plasmoids,
regardless of $\beta_0$.

\begin{figure}[htbp]
\centering
\includegraphics[width={\columnwidth}]{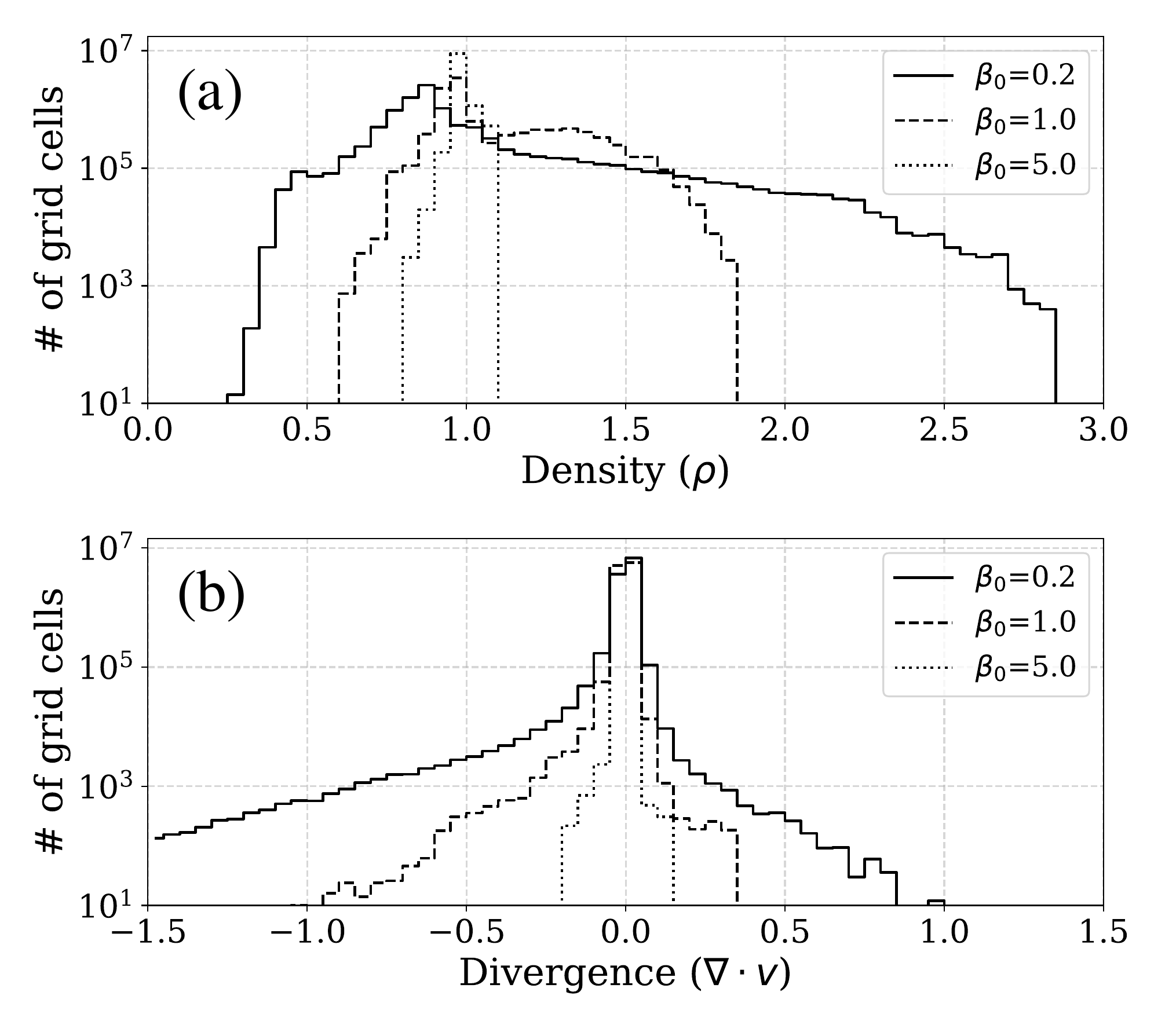}
\caption{
(a) Histogram of plasma densities $\rho$
in $1.1 \times 10^7$ grid cells in the central region
at $t=2000$ ($\beta_0=0.2$), $t=1700$ ($\beta_0=1.0$), and
$t=1800$ ($\beta_0=5.0$).
(b) Histogram of the divergence $\nabla \cdot \vec{v}$.
\label{fig:comp}}
\end{figure}

Fig.~\ref{fig:comp} shows
the distribution of grid cells in the central region,
as functions of (a) the plasma density and
(b) the divergence of the plasma velocity
at $t=2000$ for the $\beta_0=0.2$ run, $t=1700$ ($\Psi_{\rm rec} = 17.0$) for $\beta_0=1.0$, and $t=1800$ ($\Psi_{\rm rec} = 17.5$) for $\beta_0=5.0$.
The time stages are selected such that
the reconnected fluxes are similar.
In Fig.~\ref{fig:comp}(a), one can see a large variation in the density
in the high-density side for $\beta_0=0.2$ (the thick line). 
In contrast, for $\beta_0=5.0$,
plasma density is almost uniform. 
Fig.~\ref{fig:comp}(b) tells us that
plasma compression ($\nabla\cdot\vec{v}<0$) is more pronounced
than plasma expansion ($\nabla\cdot\vec{v}>0$) for $\beta_0=0.2$.
For $\beta_0=5.0$, one can hardly see signatures of compression or expansion, i.e., $\nabla\cdot\vec{v}\approx 0$.
These results tell us that
plasma compression is a key feature in the $\beta_{\rm in} < 1$ regime.

\section{A scaling model}
\label{sec:theory}

We propose that plasma compression allows faster reconnection in the $\beta_{\rm in} < 1$ regime. 
Considering that the plasmoid-dominated reconnection is an ensemble of
many mini Sweet--Parker reconnection sites, 
one can estimate the typical reconnection rate in the following way. 
When Lundquist number exceeds the critical Lundquist number,
$S_L > S_{\rm c} \approx 10^4$, the current sheet is split into
multiple Sweet--Parker layers in the plasmoid-dominated regime.
The rate of mini Sweet--Parker reconnections gives
an effective rate of the plasmoid-dominated reconnection,
$\langle\mathcal{\bar{R}}\rangle \sim S_{\rm c}^{-1/2} \sim 0.01$,
which is larger than the Sweet--Parker rate of $S_{\rm L}^{-1/2}$.
The effective rate should be proportional to
the rate of the mini Sweet--Parker layers.

We remark that the Sweet--Parker theory assumes incompressible plasmas.
The influence of the compressibility and
the inflow plasma $\beta$ ($\beta_{\rm in}$) to the Sweet--Parker theory has rarely been discussed.
\citet{hesse11b} has recently developed a compressible extension of the Sweet--Parker theory.
The authors argue that the rate of the Sweet--Parker layer should satisfy
\begin{equation}
\label{eq:hesse11R}
\mathcal{\bar{R}} \equiv \frac{E}{c_{\rm A,in}B_{\rm in}} = \frac{1}{\sqrt{2}}\sqrt{1-r (d/L)^2}
\cdot r
\Big(\frac{d}{L}\Big)
\approx
\frac{r}{\sqrt{2}}
\Big(\frac{d}{L}\Big)
,
\end{equation}
where $r=\rho_{\rm out}/\rho_{\rm in}$ is the compression factor,
$d$ and $L$ are the thickness and the length of the Sweet--Parker layer. 
The compression factor is obtained by
dropping the $\delta^2$ term in Eq.~(19) in \citet{hesse11b},
\begin{equation}
\label{eq:r_hesse}
r = \frac{\Gamma(1+\beta_{\rm in})}{3/2+\Gamma\beta_{\rm in}}
,
\end{equation}
where $\Gamma=\gamma/(\gamma-1)$. 
Aside from a minor factor, we expect that
the rate of the plasmoid-dominated reconnection scales like
\begin{equation}
\label{eq:rate}
\mathcal{\bar{R}}
\sim r (d/L)
\sim r S_{\rm c}^{-1/2}
\sim 0.01
\frac{\Gamma(1+\beta_{\rm in})}{3/2+\Gamma\beta_{\rm in}}
.
\end{equation}
The compression ratio should work as a speed-up factor in the reconnection rate.
In the incompressible limit of $\beta_{\rm in} \gg 1$,
Eq.~\eqref{eq:rate} recovers
a familiar result of $\mathcal{\bar{R}} \approx 0.01$.

We also estimate a typical composition of
the outgoing energy flux from the plasmoid-dominated reconnection site. 
Neglecting an energy flow by resistive diffusion,
the energy flux can be discussed in a form of
kinetic energy flux ($\frac{1}{2} \rho v^2\vec{v}$),
enthalpy flux (${\Gamma} p \vec{v}$), and
Poynting flux ($[-\vec{v}\times\vec{B}]\times\vec{B}$).
We assume that the typical density in the turbulent outflow,
including the plasmoids and inter-plasmoid current layers,
is comparable with the outflow density of the mini Sweet--Parker layers.
Using the pressure balance $p_{\rm out} \approx p_{\rm in} (1+1/\beta_{\rm in})$,
one can estimate an average partition of the outgoing energy flux from the reconnection site:
\begin{align}
&
\frac{1}{2} \rho_{\rm out} v_{\rm out}^3 :
\Gamma p_{\rm out} v_{\rm out} :
B^2_{\rm out} v_{\rm out}
\approx
r
\Big(\frac{v_{\rm out}}{c_{\rm A,in}}\Big)^2 :
\Gamma (1+\beta_{\rm in}) :
2 \mathcal{\bar{R}}^2
.
\label{eq:e_flux}
\end{align}
The last one, i.e., the outgoing Poynting flux is negligible, $\sim\mathcal{O}(10^{-4})$.
From the compression ratio (Eq.~\eqref{eq:r_hesse}) and
the fact $|v_{\rm out}| \approx c_{\rm A,in}$ (Fig.~\ref{fig:snapshot}(b)),
the ratio of the bulk kinetic energy flux to the enthalpy flux is
\begin{align}
1 / ( 1.5 + \Gamma\beta_{\rm in} )
\label{eq:e_ratio}
.
\end{align}

To validate these predictions,
we have carried out additional simulations.
Realizing that Eq.~\eqref{eq:r_hesse} is a function of $\beta_{\rm in}$ and $\gamma$, we have surveyed a 2-D parameter space of $(\beta_0, \gamma) \in ([ 0.2, 0.5, 1.0, 2.0, 5.0 ]$, $[ 4/3, 1.5, 5/3, 2.0 ])$. 
For $\gamma < 5/3$, more energy can be transferred to the plasma internal energy and so the compressible effects will be pronounced. 
On the other hand, plasmas are less compressible for $\gamma > 5/3$. 
For two cases of $(\beta_0, \gamma) = (0.2, 2.0)$ and $ (0.5, 2.0)$, we have added an artificial impact near the X-line at $t=1000$ so that the system switches into the turbulent state within a computation time. 
We have confirmed that this impact does not alter the properties of the turbulent state, by imposing the same impact to the $(\beta_0, \gamma) = (0.2, 5/3)$ run. As indicated by the blue curve in Fig.~\ref{fig:rate}, after the impact, the system immediately switches to the turbulent state with a similar rate. 
For the theory, we estimate $\beta_{\rm in}$ in the following way.
From the entropy conservation $p \propto \rho^{\gamma}$, we obtain a $\beta_{\rm in} = \beta_0 (1-\Psi_{\rm rec}/\Psi_0)^{\gamma-2}$.
Then we employ $\beta_{\rm in}$ for a mean flux $\Psi_{\rm rec}=12$ in our predictions. We set $\beta_{\rm in}=1.09\beta_0, 1.07\beta_0, 1.04\beta_0$, and $\beta_0$ for $\gamma=4/3, 1.5, 5/3$, and $2$, respectively.

\begin{figure}[htbp]
\centering
\includegraphics[width={\columnwidth}]{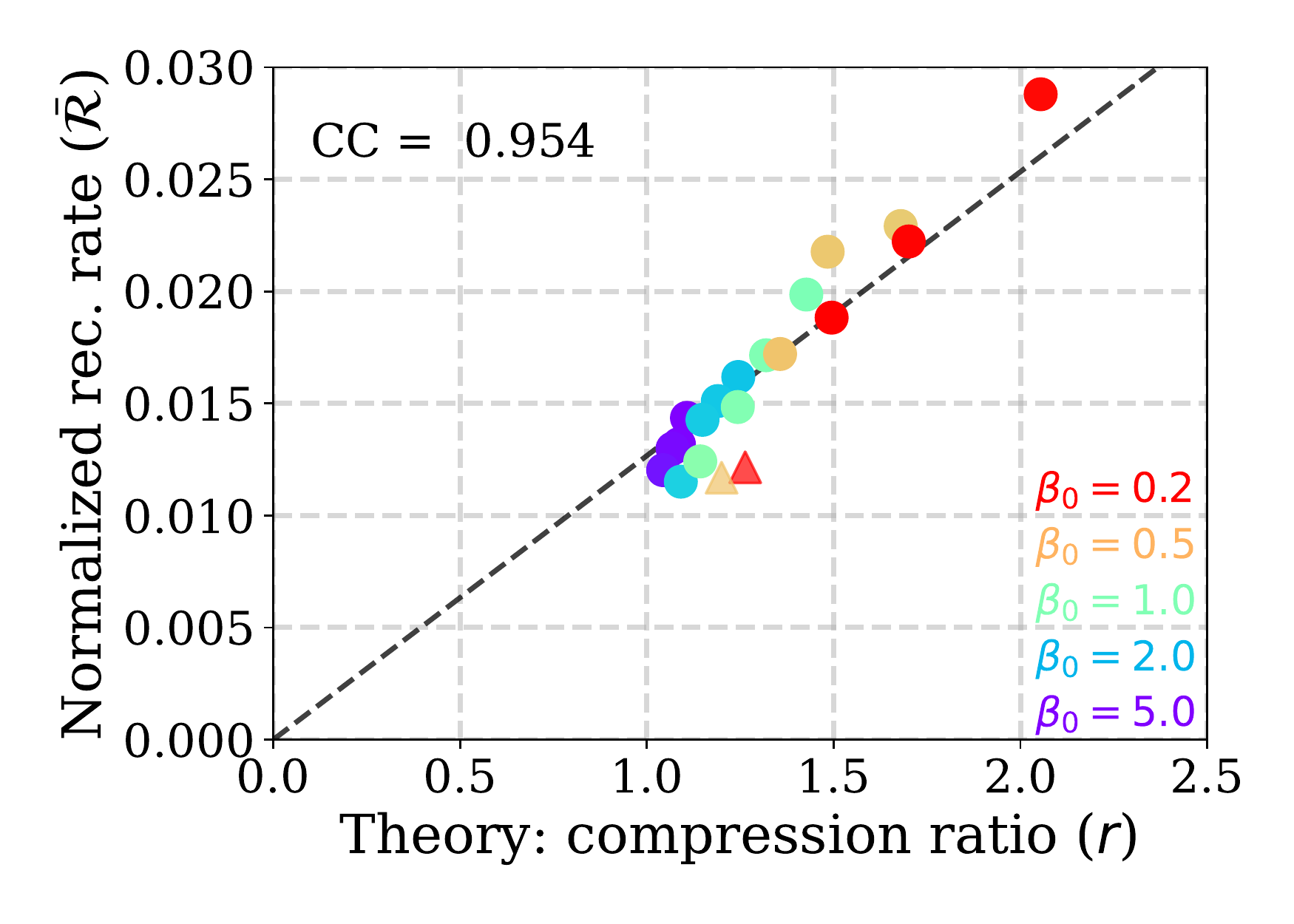}
\caption{
Reconnection rate from the simulations and
the compression ratio from Eq.~\ref{eq:r_hesse}.
The color of the circles indicates $\beta_0$.
\label{fig:theory}}
\end{figure}

Fig.~\ref{fig:theory} compares
the measured rates $\langle\mathcal{\bar{R}}\rangle$ in the plasmoid-dominated reconnection and
theoretical predictions for 20 cases. 
The color of the circles indicate $\beta_0$,
from blue ($\beta_0=5.0$) to red ($\beta_0=0.2$). 
The triangles indicate the two runs with artificial impacts. 
One can see that the measured rates are
proportional to the compression ratio.
These results are fitted by the line from the origin $(0,0)$
by the least square method.
We have obtained
\begin{equation}
\label{eq:rate2}
\langle\mathcal{\bar{R}}\rangle
\approx 0.0127
~
\frac{\Gamma(1+\beta_{\rm in})}{3/2+\Gamma\beta_{\rm in}}
.
\end{equation}
The correlation coefficient is 0.95 and
the correlation becomes even better if we drop the two triangles. 
Considering uncertainties in the theory and measurement methods,
Eqs.~\eqref{eq:rate} and \eqref{eq:rate2} are surprisingly similar.
We only recognize minor difference in the factors. 

Then we evaluate the outgoing energy flux 
during the plasmoid-dominated stage. 
In all the runs, we calculate the energy flow
at the boundary of the central domain
$x=\pm 200$ and $|y|<15$
at the time interval of $\Delta t = 10$.
By integrating them
during $t_{\Psi_{\rm rec}=7} < t < t_{\Psi_{\rm rec}=17}$,
we obtain an average profile of the outgoing energy flow.
Despite of repeated ejection of plasmoids,
the composition of the outgoing energy flux remains similar. 
Since the magnetic energy is dissipated by the reconnection process,
the outgoing Poynting flux is negligible, as estimated. 
Most of the energy is carried by the plasma energy flow,
the bulk kinetic energy flux and the enthalpy flux.

\begin{figure}[htbp]
\centering
\includegraphics[width={\columnwidth}]{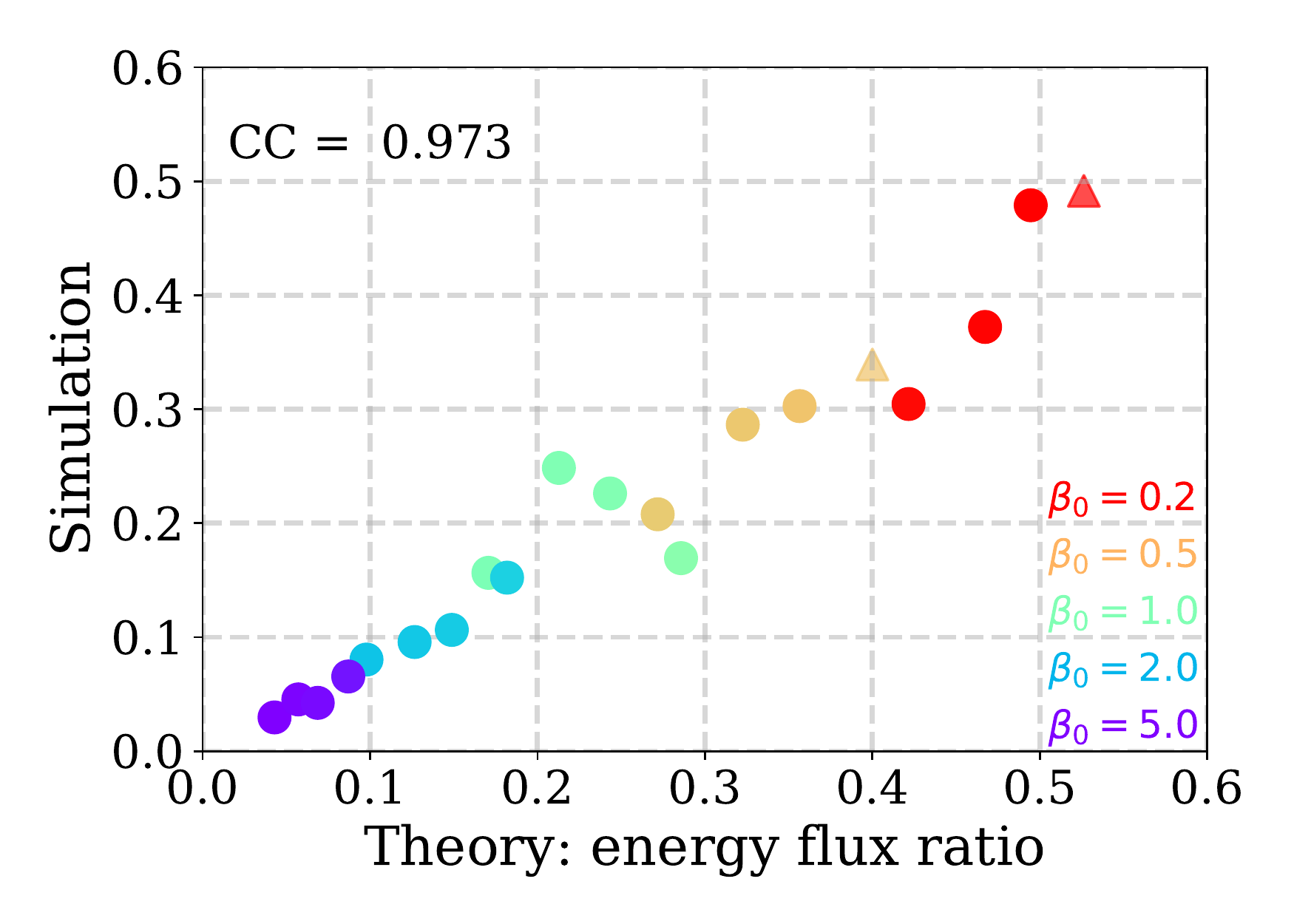}
\caption{
Average ratio of the outgoing bulk kinetic energy flux ($\frac{1}{2} \rho v^2\vec{v}$) to the outgoing enthalpy flux (${\Gamma} p \vec{v}$) from the central domain. 
Simulation results (vertical axis) and theoretical predictions (horizontal axis) are compared.
\label{fig:theory2}}
\end{figure}

Fig.~\ref{fig:theory2} shows
the ratio of the bulk kinetic energy flux ($\frac{1}{2} \rho v^2\vec{v}$) to the enthalpy flux ($\Gamma p \vec{v}$) 
during the turbulent state in the 20 runs.
Numerical results and our prediction
(Eq.~\eqref{eq:e_ratio}) are compared.
Despite variations,
one can see an excellent correlation between them.
This validates our scaling model
that contains the compression factor $r$.

\section{Discussion and Summary}
\label{sec:discussion}

In this Letter, we have studied plasmoid-dominated reconnection in low-$\beta$ background plasmas
by means of resistive MHD simulations.
To the best of our knowledge,
this is the first study to demonstrate
compressible effects in the plasmoid-dominated reconnection. 
We have found several visible signatures.
The system becomes highly complex
due to repeated formation of plasmoids and shocks. 
As evident in Fig.~\ref{fig:snapshot}(d),
many normal and oblique shocks propagate in the system. 
These shocks are successfully resolved by the shock-capturing numerical solver \citep{miyoshi05}.

We have found that the average rate increases
in the $\beta_{\rm in} < 1$ regime.
This differs from a popular statement that
the rate of plasmoid-dominated reconnection is constant $\sim 0.01$ regardless of other parameters.
We attribute this to compressible effects. 
Recognizing that the plasmoid-dominated reconnection consists of many mini Sweet--Parker layers,
we have proposed a simple scaling model for the reconnection rate (Eq.~\eqref{eq:rate}),
based on the compressible Sweet--Parker theory \citep{hesse11b}. 
The rate is accelerated by a speed-up factor of Eq.~\eqref{eq:r_hesse}.
We have tested our predictions
in the 2-D parameter space of $(\beta_0, \gamma)$. 
The numerical results are in good agreement with predictions,
both in the reconnection rate and in the energy flow.

In a solar corona with $\beta \ll 1$,
Eq.~\eqref{eq:rate2} tells us that
the reconnection rate increases by a factor of 5/3 and that
it approaches $0.02$. 
The numerical survey also indicates that
the plasmoid-dominated reconnection can be faster
for smaller adiabatic index, $\gamma < 5/3$.
If heat conduction, viscosity, radiation, or other effects
enhance the effective plasma compressibility,
we expect even faster rate of reconnection.
Theoretically, the rate of plasmoid-dominated reconnection determines a lower bound of the energy release rate by magnetic reconnection, and therefore
a solar flare can be a twice faster energy converter
than previously thought.

As mentioned, \citet{ni12,ni13} and \citet{baty14} studied the onset of the plasmoid-dominated reconnection in a low-$\beta$ background plasma. 
They observed that reconnection becomes turbulent earlier for larger $\beta_0$, however, it is difficult to compare our results to theirs, because we have triggered the turbulent state in a very different way. 
\citet{ni12,ni13} further reported
higher reconnection rates for higher $\beta_0$.
This was attributed to the initial density variation. 
In this study, we have studied later stages that are virtually unaffected by the initial profiles. 
In fact, magnetic fluxes in the outflow region and in the initial current sheet are estimated to be $\Psi_{\rm out} = \mathcal{R}  L_x/2$ and $\Psi_{\rm cs} = 1$.
When the reconnection proceeds at the rate of $\mathcal{R} \le 0.02$, the current-sheet plasmas were lost from the reconnection region at the time of $\Psi_{\rm rec} = \Psi_{\rm out} + \Psi_{\rm cs} \le 6$ in all cases. 
We started our measurement at the time of $\Psi_{\rm rec}=7$,
after the system lost the memories of the initial profile.

Our results have implications for
magnetic reconnection in astrophysical settings,
where radiative cooling and relativistic fluid effects are important. 
\citet{uzdensky11} have envisioned that
intense radiative cooling leads to a higher reconnection rate,
because of plasma compression. 
Our results are favorable to their argument, because
we demonstrate that the plasma compression allows faster reconnection. 
\citet{takamoto13} has studied the plasmoid-dominated reconnection in a relativistic plasma, and has reported an average rate of ${\sim}0.02$ for $\beta_0=0.2$.
This rate is twice faster than a typical nonrelativistic value.
This is possibly due to the enhanced compression in a relativistically hot plasma with the adiabatic index $\gamma=4/3$. 
Although our model is nonrelativistic, Eq.~\ref{eq:r_hesse} suggests substantially higher compression for $\gamma=4/3$.


We note that we have explored a basic configuration with antiparallel magnetic fields. Reconnection may occur in skewed configurations with an out-of-plane background field (``guide field'') $B_z$. 
The guide field modifies $\beta_{\rm in}$ but provides a magnetic pressure that resists compression.
In fact, equations for $B_z$ are similar to fluid equations with $\gamma = 2$, and our results for $\gamma = 2$ are more conservative than for $\gamma = 5/3$. 
Thus the compressible effects should be less pronounced
in the presence of a strong guide field. 
Unfortunately, we cannot make a more quantitative prediction at this point,
because our underlying theory only covers the antiparallel case.
Further theoretical and numerical investigations are necessary for situations with a guide-field. 

The physics of magnetic reconnection has been organized by
the Lundquist number and the thickness of the reconnection layer \citep{huang11,ji11}. 
The latter parameter corresponds to the collisionality,
and its role reminds us the Knusen number in fluid dynamics. 
This study further extends these understandings.
We propose that the inflow $\beta$ ($\beta_{\rm in}$) and relevant parameters ($\gamma$ and $\mathcal{M}_s=[\frac{1}{2}\gamma\beta_{\rm in}]^{-1/2}$) characterize the physics of magnetic reconnection
at least in an antiparallel case.
In the low $\beta_{\rm in}$ (high $\mathcal{M}_s$) regime,
compressible effects become prominent ---
the system is shock-dominated and the compression allow faster flux transport.

\begin{acknowledgments}
Simulations were carried out
on facilities at Center for Computational Astrophysics,
National Astronomical Observatory of Japan,
on the JSS2 system at Japan Aerospace Exploration Agency, and
on the A-KDK system at Kyoto University.
This work was supported by Grant-in-Aid for Scientific Research (C)17K05673
from the Japan Society for the Promotion of Science (JSPS).
\end{acknowledgments}

\end{document}